\documentclass{aastex701}

\usepackage{booktabs}
\usepackage{amsmath}

\begin{document}

\title{Towards millimagnitude Photometry at the Vera Rubin Observatory: \\
Aerosol Monitoring with Quadband Dispersed Imaging.}

\author[0000-0002-8883-2172]{Eske M. Pedersen}
\affiliation{Department of Physics, Harvard University \\
17 Oxford Street, Cambridge, MA, USA
}
\email{eskepedersen@fas.harvard.edu}
\author[0000-0003-0347-1724]{Christopher W. Stubbs} 
\affiliation{Department of Physics, Harvard University \\
17 Oxford Street, Cambridge, MA, USA
}
\email{stubbs@g.harvard.edu}
\affiliation{Department of Astronomy, Harvard University}
\author[0000-0001-9440-8960]{Merlin Fisher-Levine} 
\affiliation{Department of Astrophysical Sciences \\
Princeton University \\ 
}
\email{merlin.fisherlevine@gmail.com}
\affiliation{Vera C. Rubin Observatory, La Serena, Chile}

\author[0000-0002-3205-2484]{Elana K. Urbach}
\affiliation{Department of Physics, Harvard University \\
17 Oxford street, Cambridge, MA, USA 
}
\affiliation{Vera C. Rubin Observatory, La Serena, Chile}
\email{eurbach@g.harvard.edu}

\author[0000-0003-2852-268X]{Erik Dennihy}
\affiliation{Vera C.\ Rubin Observatory Project Office, 950 N.\ Cherry Ave., Tucson, AZ  85719, USA}
\email{edennihy@lsst.org}

\author{Patrick Ingraham}
\affiliation{Steward Observatory, University of Arizona, 933N Cherry Avenue, Tucson, AZ, 85721,
USA}
\email{pingraham@arizona.edu}

\begin{abstract}



As the frontier of precision astronomical photometry continues to advance, correcting for time-variable atmospheric transmission becomes increasingly important. We describe an observational approach to monitoring optical attenuation due to atmospheric aerosols, using a multiband filter and disperser on the Auxiliary Telescope at the Vera C. Rubin Observatory. This configuration allows us to perform simple aperture photometry on four notched-out spectral regions, covering 347 to 618 nm. We see clear evidence of temporal variations in extinction across these bands, which we attribute to variation in the aerosol content of the atmosphere above the observatory. The observed {\it differences} in extinction between the reddest and bluest band can exceed 5 mmag/airmass, highlighting the importance of including variable aerosols in the transmission of the atmosphere. We aspire to using precise determinations of the optical transmission of the atmosphere to enable a forward-modeling approach to achieving mmag photometric precision with Rubin data.   

\end{abstract}


\section{Introduction} \label{sec:intro}

The Vera C. Rubin Observatory is poised to become one of the world's premier sky survey systems. With a science reach that spans from the solar system to cosmology  
\citep{ivezic2019lsst}, Rubin's combination of unprecedented etendue, fast focal plane readout, and real-time data dissemination will serve as a foundational resource for the astronomical community in the decades ahead.  

One of the primary science goals of the Rubin system is to better elucidate the nature of the Dark Energy that is driving an accelerating cosmic expansion
\citep{riess1998observational,perlmutter1999measurements}. A range of techniques will be exploited to analyze the Rubin images to better constrain cosmological parameters, including strong, weak, and micro lensing, cluster abundance studies, and 3-d mapping of the large scale structure. The direct measurement of the history of cosmic expansion using type Ia supernovae, along different lines of sight, is another powerful probe of cosmology. With the number of type Ia SNe we anticipate being detected by Rubin, there is an urgent need to improve our photometric precision \citep{stubbs2015precise}. As emphasized in \cite{stubbs2006toward}, making explicit measurements of the wavelength-dependence of both atmospheric transmission and relative instrumental throughput can alleviate the intrinsic ambiguity of using broadband photometry of celestial spectrophotometric calibrators as the primary metrology foundation for supernova photometry. 

In anticipation of the need to characterize the time-variable aspects of atmospheric transmission that have historically impeded our ability to achieve millimagnitude levels of photometric precision, the Rubin observatory includes a dedicated 1.2m Auxiliary Telescope that uses stars to back-light the atmosphere. By performing spectrophotometric measurements of a network of appropriately bright stars, the time-variable aspects of atmospheric transmission, namely aerosols, ozone, and water vapor \citep{stubbs2007toward}, can be determined with sufficient precision to make appropriate photometric corrections. Ideally, this would entail forward-modeling \citep{burke2017forward} using the appropriate photon spectral distribution for a supernova of a given type and redshift, at a known time past detonation. Knowing both the atmospheric and instrumental optical throughput functions vs. wavelength is essential in order to carry out this process. 

Our overall strategy is to use atmospheric transmission modeling such as MODTRAN \citep{modtran6} for the deterministic aspects of optical transmission (Rayleigh scattering and O2 absorption). This will be supplemented with adjustments for water vapor, ozone, and aerosol content. The resulting atmospheric transmission function can be used in conjunction with explicit measurements of the Rubin instrumental transmission function and of Galactic extinction to perform the forward modeling described above. Early tests were done using the CTIO 0.9m telescope \citep{coughlin2018testing,neveu2023slitless} and we are now implementing this scheme at the Rubin site on Cerro Pachón. Having a telescope dedicated to the characterization of the atmosphere relieves the tension between measuring extinction using standards and carrying out a program of science observations. Our objectives are similar to those described in \cite{patat2011optical}, but with the aim of attaining mmag levels of precision. 

The need to move beyond the conventional approach of using broadband photometric measurements at varying airmass to estimate extinction, under the assumption that atmospheric attenuation only depends on zenith angle, is outlined in \cite{mcgraw2009measurement}, \cite{louedec2015atmospheric}, and \cite{li2016assessment}. A corresponding level of detail must be brought to bear in suppressing instrumental sources of measurement error, as described in \cite{bernstein2017instrumental} and related papers.  The angular and temporal correlation functions of the variable components of atmospheric attenuation are a topic of ongoing investigation, building on prior work \cite{li2017temporal} and \cite{burke2010precision,burke2013all}. The subtleties of achieving atmospheric corrections that support photometry at the millimagnitude level were thoughtfully discussed in the final stages of the photoelectric photometry era. We note in particular the in-depth paper by \cite{young1991precise} that reviews the need to properly account for the interplay between source spectrum and atmospheric attenuation.  This move towards millimagnitude precision was derailed by the advent of CCDs. The manifest advantages of silicon CCDs led to their rapid adoption by the astronomical community, but cosmetic defects and small array sizes in early CCD generations were a setback to precision photometry. Over the past decade, the astronomical community has clawed its way back to the millimagnitude photometric precision frontier. 

The reader might well argue that measurements of planetary transits, using differential photometry of stars seen in successive images, has gone far beyond millimagnitude precision. That's true, but the state-of-the-art in linking photometric measurements across all-sky survey data does not yet achieve this performance. This difference in precision between differential photometry (taken at the same time through the same atmosphere) and observations that are separated in time and angle suggests that variable atmospheric attenuation is a major limiting factor in attaining high-precision survey photometry.

More formally, the flux $\Phi$ that is detected from some celestial source is \citep{stubbs2007toward}

\begin{equation}
   \Phi = \int SPS(\lambda) \cdot T_{\text{EG}}(\lambda) \cdot T_{\text{Gal}}(\lambda) \cdot T_{\text{atmos}}(az, alt, t, \lambda) \cdot T_{\text{inst}}(\lambda,t) d\lambda,  
\end{equation}

\noindent
where $SPS$ is the source photon spectrum, $T_{\text{EG}}$ is transmission through extragalactic extinction, $T_{\text{Gal}}$ is transmission through Galactic extinction along the line of sight, $T_{\text{atmos}}$ is transmission through the Earth's atmosphere, and we have listed explicitly its dependence on azimuth, altitude, and time, and $T_{\text{instr}}$ is the instrumental throughput, including detector quantum efficiency.

Our focus here is atmospheric transmission, $T_{\text{atmos}}(az, alt, t, \lambda)$. The conventional assumption is that during photometric nights this quantity depends only on airmass, {\it i.e.} telescope altitude. But as we push the limits of photometric precision into the millimagnitude regime we need to pay careful attention to the angular and temporal correlation functions of the non-deterministic wavelength-dependent aspects of atmospheric attenuation, namely water vapor, aerosols, and ozone.   

There are two distinct conceptual approaches to using stars to backlight the atmosphere and deduce its optical attenuation. If the photon spectrum of the source and the instrumental throughput are both fully known, then a single spectrum suffices to determine the optical attenuation vs.~wavelength along that sightline. After appropriate flat-fielding, obtaining a wavelength solution, and accounting for spectral blending due to variable PSF effects, dividing the observed spectrum by the product of stellar photon spectrum and the instrumental throughput function yields a quotient that is the atmospheric transmission function. An alternative approach is to exploit the airmass dependence of atmospheric attenuation and extract the spectral components that are airmass-dependent. This requires no prior knowledge of source spectra or instrumental properties, but does presume that both of them are temporally stable over the course of the observations. It also assumes that any temporal variations or azimuth-dependence are well-sampled and can be separated from the zenith-angle dependence.  We are in the process of exploring what observing strategy will yield the best estimates of the variable components of atmospheric transmission. 

This paper describes initial results from a method that combines a dispersive element with a multiband filter to produce spectra that can be analyzed with photometric rather than spectroscopic techniques. This avoids issues with second-order light contamination and the need to obtain precise wavelength solutions. 
We begin by summarizing work undertaken in prior sky surveys for the determination of atmospheric extinction. We then briefly describe the Rubin Auxiliary Telescope, its instrument, and the data it provides. This is followed by a description of the analysis methodology for dispersed multiband imaging. We then assess the time evolution of the inferred aerosol extinction. We highlight some of the shortcomings we find, and how they will be addressed. 

\subsection{Prior examples of atmospheric characterization for photometric surveys}

The Rubin project is not the first attempt at supplementing a broadband sky survey system with a smaller-aperture telescope that is tasked with the determination of atmospheric extinction. The SDSS survey set up a photometric monitoring telescope \citep{tucker2006sloan} that used a filter wheel to obtain sequential images in passbands designed to mimic those of the main survey. The DES survey took the approach of obtaining simultaneous images \citep{li2012atmcam,li2014monitoring} in multiple bands, some of which were designed to determine water vapor content. 
An alternative approach for measuring water vapor is to exploit differential dispersion at radio frequencies, using dual-band GPS receivers \citep{blake2011measuring, wood2022gps}. Spectrophotometric measurements have been done previously \citep{burke2010precision, burke2013all, coughlin2018testing, neveu2023slitless}, but these were short observing campaigns, not done on an ongoing basis. A comprehensive analysis of ongoing spectrophotometric atmospheric characterization at Manuna Kea is presented in \cite{buton2013atmospheric}.  

Downward-looking remote sensing data can contribute to atmospheric characterization above an observatory, but the revisit cadence poses a challenge. This may change in the future with the increase of low-altitude constellations, and perhaps establishing emitters on orbit \citep{peretz2021orcas}. In the near term, determining the ozone content seems the most likely use of remote sensing data \citep{guyonnet2019local}. 

\subsection{Atmospheric Aerosols}
\label{sec:aerosols}

Aerosols arguably pose the most challenging aspect of correcting our photometry for variable atmospheric optical attenuation. The particulates that scatter light out of the beam can have a time-variable optical depth, as well as varying size and shape distributions.  The gray (wavelength-independent) aerosol component has the same effect as clouds, and given enough repeated observations the methods of multi-epoch photometric calibration can address this aspect. The wavelength-dependent component of aerosol scattering is more subtle and depends on both the size and shape distribution of the particulates. A common parameterization of optical depth due to aerosols is:
\begin{equation}
\label{eq:optical_depth}
    \tau (\lambda) = \tau _ {500} (500 nm /\lambda)^{\alpha}.
\end{equation} 
We will adopt this formulation where the Angstrom exponent $\alpha$ captures the wavelength dependence and $\tau_{1 \mu m}$ the scattering strength. At a monochromatic wavelength $\lambda_0$ the relationship between optical depth and extinction coefficient $dm/d\chi$ (in magnitudes per airmass) amounts to $dm/d\chi=1.0857 \tau(\lambda_0)$.  

 A substantial data set on atmospheric aerosols has been accumulated by the AERONET system \citep{holben1998aeronet}. The AERONET system uses multiband observations of scattered sunlight across a variety of scattering angles to deduce the aerosol burden above a fixed site. See \cite{stubbs2007toward} for observational evidence from AERONET for substantial variation in $\alpha$ at Mauna Loa. The spectrophotometic analysis for Mauna Kea that is presented in \cite{buton2013atmospheric} concludes that aerosols contribute their dominant source of variation and uncertainty in extinction determination at optical wavelengths.  The Burton {\it et. al.} measurements of the optical depth due to aerosols can often rise up to a few percent, with substantial variation (or uncertainty) in the Angstrom exponent.  A comparison of multiband photometry with MODTRAN extinction predictions \cite{stubbs2015comparison} showed immeasurably low aerosol attenuation above Mt. Hopkins, on a single night. A long-term aerosol monitoring campaign at a continental European site is described in \cite{reimann1992atmospheric}. They show that the Angstrom exponent can have large excursions, depending on the physical characteristics of the scattering particulates. 

An alternative method to measure atmospheric aerosol content is to send light from a short-pulse laser up into the atmosphere and measure the intensity of the backscattered light. This LIDAR method has the advantage of providing a vertical profile of the scattering material. The challenge in interpretation is converting from backscatter cross section to an estimate of optical depth and its wavelength dependence. This approach is described by \cite{dawsey2006lidar, zimmer2006ale} and \cite{zimmer2012new}. Our approach is to measure directly the line of sight optical attenuation at the wavelengths of interest, by backlighting the atmosphere with stars. 

\section{Methodology} \label{sec:method}

\subsection{Rubin's Auxiliary Telescope and the LATISS instrument.} \label{sec:Auxtel}

A description of the calibration hardware, including the Auxiliary Telescope, for Rubin is provided in \cite{ingraham2022vera}. Briefly, the Auxiliary Telescope is a 1.2m Ritchey-Chretien telescope that feeds a beam to one of two Nasmyth foci. The LATISS instrument \citep{mondrik2020calibration} has a filter wheel and a disperser wheel that can introduce various combinations of optical elements into the f/18 converging beam.
With a beam this slow, at the modest resolutions we use for atmospheric characterization, we can place dispersing elements into the beam without the spatialmask-collimation-dispersion-reimaging configuration that is traditional for astronomical spectrographs \footnote{We thank David Monet for this suggestion, based on his experience at USNO.}. The CCD in the LATISS instrument is a 4K x 4K deep-depletion array with 10 micron pixels, and is one of the devices fabricated by ITL \citep{lesser2017results} for the main Rubin camera system. This CCD has 16 readout amplifiers, each configured as 512 x 2000 pixels. The readout electronics uses the same highly parallelized scheme as the main Rubin camera \citep{gilmore2008lsst} and this allows us to read out the array in 2 seconds.  Typical read noise at this pixel rate is 8 electrons rms. The plate scale of the LATISS instrument is 0.1 arcsec per 10 $\mu$m pixel. 

Our scheme uses masked dispersed imaging to perform spectrophotometry without being afflicted by variation in slit losses due to differential atmospheric refraction. In order to minimize ghosts the LATISS instrument does not have an atmospheric dispersion corrector (ADC). This configuration has the added advantage that regardless of the spectrum of the night sky, each pixel sees multiple wavelengths that are illuminating it from different angles of arrival at the pupil. As a result, despite being a dispersed system the sky background is flat on the detector and this greatly simplifies sky subtraction. We typically obtain good signal to noise ratios on stars as faint as 9th magnitude in 30 seconds of integration. The spectra are aligned along the long direction of the amplifier segments. 

The masked dispersed images are always acquired with the dispersion direction parallel to the horizon. The effect of atmospheric chromatic refraction is to bend rather than compress or expand the spectrum. This eases the task of obtaining a wavelength solution since the dispersion is then independent of zenith angle. Pointing errors do introduce a displacement of the spectrum which must be measured and corrected for.    
We have experimented with a variety of dispersers. This includes Ronchi gratings, which by construction suppress second order light and have a flat response with wavelength, and custom holographic dispersers \citep{moniez2021transmission, neveu2023slitless}. Light of different orders (m=0, $\pm$1, $\pm$2) are transmitted by these dispersers, to varying degrees. We have the ability, with the filter wheel, to place second-order blocking filters in the beam to minimize contamination in the spectral region $\lambda >$800 nm where we measure the equivalent width of water vapor absorption lines.  

We used a multiband filter in conjunction with a blazed grating (mounted in a common frame) in a configuration we call the `quadband disperser.' This is described in more detail below. 

During the construction phase of the Vera C. Rubin Observatory the collection of data with the Auxiliary Telescope has been in competition with other demands of the project. We have nevertheless obtained data for a few nights per month over the past few years. The quadband disperser was installed on Feburary 14th 2023. We present here the analysis of initial data from the system, that demonstrates both its capabilities and limitations.    

\subsection{Representative Images and Spectra}

\begin{figure}[ht]
    \centering
    \includegraphics[width=7in]{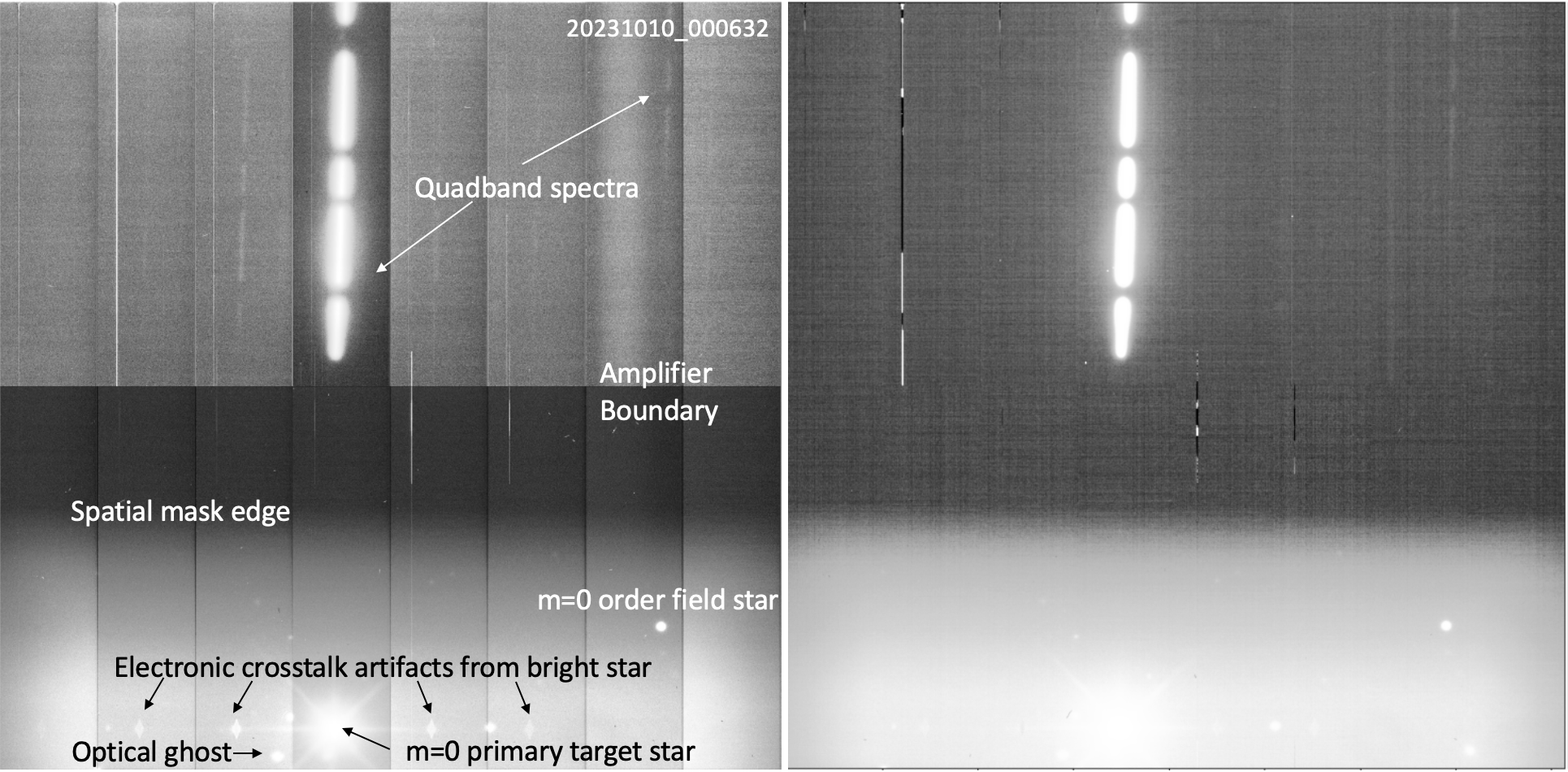}
    \caption{The panels show masked, filtered, dispersed images of HD 2811 before (left) and after (right) instrument signature removal. The sixteen readout amplifers are clearly distinguishable in the left hand panel. The m=0 order images of stars appear in the unmasked lower region of the image. Their corresponding quadband filtered m=1 spectra appear in the upper half. We perform aperture photometry on the individual spectral streaks for images obtained at different airmasses, to determine extinction in each sub-band. The spatial mask suppresses any possible spectral contamination from superimposed m=0 field stars, Note that (by design) then entire spectrum of interest lands on a single amplifier. The clear gaps between the passbands greatly facilitate data reduction, allowing aperture photometry of the four distinct spectral regions. The spectral band spilling off the top of the sensor is not used. The gray-scale stretch on the left hand image is computed on a per-amplifier basis.}
    \label{fig:FigISR}
\end{figure}

Figure \ref{fig:FigISR} shows an example of a masked, dispersed image obtained on the Rubin Auxiliary Telescope. The image of HD2811 (a K0IV star with V=9.4 mag) shown in Figure \ref{fig:FigISR} was taken on MJD 60228. In this 60 sec integration we obtain between 50K and 250K ADUs per column-summed spectral bin.  

\begin{figure}[ht]
    \centering
    \includegraphics[width=6.0in]{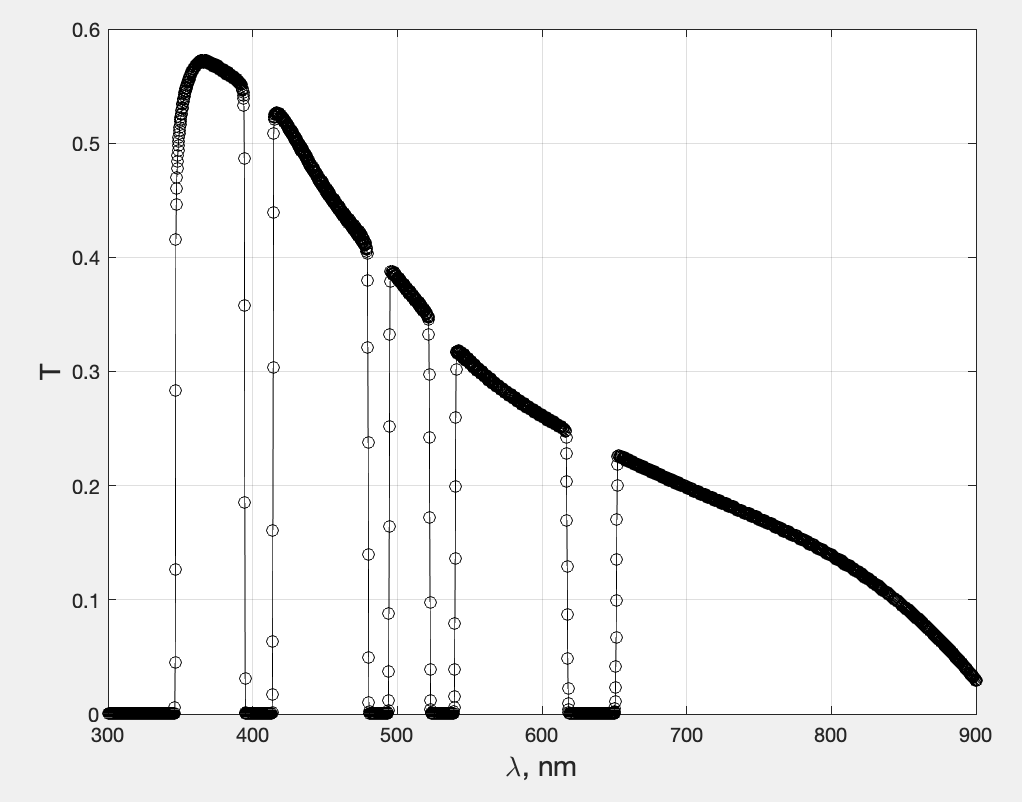}
    \caption{This figure shows the throughput of the combined grating plus multiband interference filter. The system is configured so that the spectral region $\lambda >$ 650 nm, part of which is contaminated by second order light, falls off the edge of the detector. We use the bluest four spectral segments, designated as bands B1-B4.  }
    \label{fig:quadgrating}
\end{figure}

Our quad-band approach inserts a sequence of a spatial mask, a multiband interference filter (Semrock NF03-405-488-532-635E-50x50), and a blue-blazed 300 line/mm transmission grating (Edmund Optics part number 85-296) into the beam. The transmission characteristics of this combined optical element (including a very small perturbation averaged over the modest angles of incidence) is shown in Figure \ref{fig:quadgrating}. This configuration has the advantage of placing the entire quadband spectral region of interest onto a single amplifier, rendering us less sensitive to systematics due to gain-matching errors across the CCD. The beam footprint at the grating from a point source is an annulus with a diameter of 13mm. The 50mm x 50mm disperser+filter pair is offset from the optical axis of the system to optimize the spectral footprint on the sensor.  We fit a dispersion of 0.177 nm/pixel. 
The interference filter is placed ahead of the disperser, to minimize the range of incidence angles to the f/18 Aux Tel beam. 

The blue and red edges of the multiband filter, accounting for the angular spread of the incident f/18 beam, are listed in Table \ref{tab:edges}.

\begin{table}[ht]
    \centering
    \begin{tabular}{|c|c|c|}
    \hline
    Band & Blue Edge (nm) & Red Edge (nm) \\
    \hline
      B1 & 347.0 & 394.8 \\
      B2 & 414.4 & 480.3\\
      B3 & 494.9 & 522.9 \\
      B4 & 540.5 & 618.0\\
      \hline
    \end{tabular}
    \caption{This table lists the 50\% of peak transmission filter edge wavelengths for the QuadBand filter, after accounting for the f/18 incident beam. }
    \label{tab:edges}
\end{table}

The spatial mask can be considered as a slit with a width of 15mm in the spectral direction and extending over the full 50mm in the spatial dimension, placed at the location of the out-of-focus disperser wheel. The slit width is considerably larger than the size of the pupil image at that location. This ensures that there are no slit losses due to atmospheric refraction. The mask suppresses any zeroth order (m=0) light from field stars landing atop the m=1 spectrum of the target star. 


Some advantages of the mask + quadband + disperser configuration are 
\begin{itemize}
\item We can use aperture photometry rather than spectroscopic analysis methods, with no need to obtain detailed frame-by-frame wavelength solutions. The edges of the interference filter passbands clearly define the spectral regions over which we integrate flux.   
\item The data are insensitive to changes in focus or seeing. Poor seeing softens the slope of the sub-band edges but the illuminated regions remain well-separated. 
\item The sky background is greatly reduced, for three reasons. First, since the total optical bandwidth is lower than using an unfiltered disperser we only capture a subset of the sky photons. Second, the spatial mask blocks the m=0 sky background in the region of the spectrum. The only sky photons that illuminate the upper half of the detector are dispersed. 
\item The residual sky spectrum is devoid of any emission line features. In our spectral region of interest there are very few emission lines, and they are convolved with the width of the spatial mask. This makes sky subtraction very straightforward. 
\end{itemize}

\subsection{Observations}
\label{sec:observations}

In this paper we restrict our analysis to Aux Tel observations of a single target star, HD2811, taken over 30 nights in the period from June 2024 to October 2024. This resulted in a total of 536 exposures.  By focusing on a single star we avoid having to make color-airmass corrections that are required to fuse extinction data from multiple stars on multiple (variable) nights. This object is a Calspec standard star \citep{CalSpec2020}. HD 2811, is an A3V star with B=7.67 and V=7.50 (Vega-based magnitudes) located at $\alpha$=00:31:18.4899 $\delta$=-43:36:22.998 (J2000) \cite{wenger2000simbad}. This object transits a wide range of airmass as seen from Cerro Pachon. The blue spectrum of this A star is well-matched to the Aux Tel system; we achieve high signal-to-noise ratios in 30 second quadband exposures across the wavelength range of interest. For each pointing, we obtained triplets of images in which the spectra were placed on three distinct amplifiers on the CCD. This redundancy allows us to constrain and quantify instrumental systematic errors. 

\subsection{Photometry} \label{sec:photometry}
Our photometric analysis of the dispersed quadband images entailed the following steps: First we perform a basic instrument signature removal (ISR) on the image, this includes removing known defects and a row-by-row overscan correction; for this we use the LSST pipeline's built in ISR removal. Next we conduct an object detection to find the m=0 (undispersed) image to the target star. For this we rely on the LSST Pipeline's threshold based object detection routine. The images are set up such that the undispersed target star should land on the lower half of the image, while the quadband spectra should fall on the amplifier above it. We check if the star's position on the image aligns with the expected quadband position according to the metadata. We then pass the cutout of the amplifier on to a more detailed aperture detection. 

We note that we did not perform any multiplicative ``flatfield'' correction to the images. The detector sees the convolution of the dust spots on the dewar window with the flux distribution of the dispersed light as it passes through the window. This smears out the annular dust spots we're accustomed to seeing in undispersed images. A complete flat-fielding process for LATISS is under development. We stress that we are making {\it differential} measurements of atmospheric attenuation. As long as the target star is placed at the same location on the imager, we are insensitive to flat-field issues. The triplets of images we acquired, using three different amplifiers on the CCD, sets a limit on the systematic errors that might be imposed by dust spots in the beam. 

We redo the object detection but this time just on the cutout of the amplifier. If the detection routine doesn't detect any quadbands or detects fewer than 4 objects on the amplifier we reject the image and flag it either with no detection or for issues with separating the bands. In the case of no detection, we might want to check if we accidentally pulled the wrong amplifier while in the separation case it could be an issue of image quality meaning we can't safely distinguish the bands from each other. Next we also check if any of the detected objects centroid lies too close to the upper edge of the amplifier. This is done to avoid potentially reading the second order band as one of the quad bands we are interested in, or if the bands didn't land properly on the amplifier. If after excluding any problematic bands we are  left with fewer than the 4 bands we expect, we will also exclude and flag the image. 

Next we generate the edges in the y direction from the initially detected bounding boxes. Rather than relying on the bounding boxes, we define the y-edges such that in between the 4 bands there is no space not assigned to one of our new aperture boxes. This is done by defining the y-edges as the midpoint between the lower edge of one of the initial boxes and the upper edge of the initial box below it. For the lowest aperture box, we take the lower edge of the initial box and extend it\footnote{In the code an optional parameter \texttt{yWindow} indicates how much we extend either outer edge, though on the lower edge is extended twice this number of pixels.} to ensure we are catching all the flux of the first band. Similarly for the upper aperture box we extend from the upper edge of the initial box. If either the lowest edge or the highest edge exceeds the amplifier we flag this as well. 

With the aperture edges defined in the y direction we turn to the x direction. We first find the midpoint between the y-edges of the aperture. Using 10 rows across the amplifier around this mid point we take the median across the rows, this gives us cross-section across the amplifier for each aperture box. To the cross-section we then fit a double Gaussian function as shown in Equation (\ref{eq:doubleGauss}). 
\begin{equation}
\label{eq:doubleGauss}
f(x) = a\times\exp\left[-\frac{\left(x-\bar{x}\right)^2}{2\sigma_1^2}\right] + \left(1-a\right)\times\exp\left[-\frac{\left(x-\bar{x}\right)^2}{2\sigma_2^2}\right].
\end{equation}
This gives us the center along the x-axis of the Quadband ($\bar{x}$), along with an indication of the spread $\sigma_1, \sigma_2$. Next we could either try and use the spread to dynamically estimate the bounds of the apertures or we can provide a predetermined extend in pixels. We chose the latter for consistency, by providing the optional parameter \texttt{xWindow=200}, this means the aperture will be defined as 100 pixels (10 arcsec) on either side of the $\bar{x}$. 

With the full aperture box defined we next subtract a background from the flux in each aperture. This is done by taking 10 columns on either side of the aperture, and determining a robust median across these columns for each row, this median will be used as our background and that we subtract from the aperture. 
The standard deviation in these background columns also serves as an estimate on the per pixel flux uncertainty. 

Finally we sum up all the flux for each aperture box. Besides the total flux in each aperture we also save the mean of the background columns. Along with this we count the number of pixels in each aperture that have fluxes above a previously defined threshold (for this paper we used \texttt{threshold=100000}). This together with the pixel flux percentiles from the 60th to the 100th forms our detection of over saturation.

If any of the 4 bands in one exposure had a pixel above a defined saturation threshold, we flagged the entire exposure and excluded it from further analysis. 

\begin{figure}
    \centering
    \includegraphics[width=\textwidth]{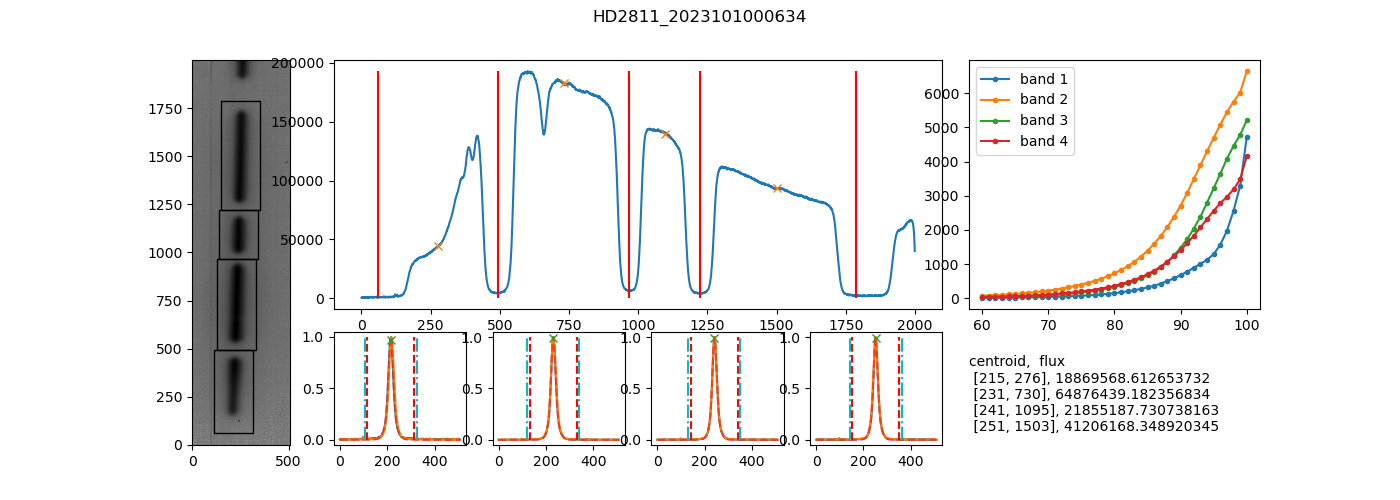}
    \caption{An example of the diagnostics plots produced by the flux estimation code. The left panels shows the amplifier where the band is detected, center top we have the cross-sum along the y-direction of the amplifier, with the red lines designating the y-bounds of the different bands. Also marked are the center-y positions. We note that we can clearly see the H-$\gamma$ feature in the stellar atmosphere in apparent at x$\sim$600, with other stellar Balmer lines seen in the bluer band. The bottom shows a cross-section along the x-direction for each band. Also plotted is the double-Gaussian fit, and the boundaries of the aperture box (red-dashed) along with the boundaries of the background box (blue-dashed). The box on the right shows the percentiles of the pixel fluxes for each of the four bands. This particular plot is for HD2811 and is exposure number 634 taken on October 10th 2023.}
    \label{fig:example_diag}
\end{figure}

We show an example of a single image's process in Figure \ref{fig:example_diag}, on the left side we have the full amplifier with the aperture boxes drawn in. The upper part of the center is a cross section along the y-direction, while each aperture's x-direction cross-section is displayed below. Finally, on the right side we plot out the pixel flux percentiles for each of the 4 apertures, this last plot can be used as a diagnostic for determining if we have pixels that are over-saturated. 

We converted the fluxes for each band to instrumental magnitudes:
\begin{equation}
    m_{\text{inst}} = -2.5\text{log} \left(\frac{f}{t_{\text{exp}}}\right),
\end{equation}
where $m_{\text{inst}}$ is the instrumental magnitude, $f$ is the flux of the band and $t_{\text{exp}}$ is the exposure time. 

The resulting table of times, four instrumental magnitudes, and associated airmasses comprise the data we used for atmospheric extinction analysis, described in the next section. 

\subsection{Atmospheric Extinction Determination} \label{sec:extinction}

We are primarily interested in measuring temporal changes in wavelength-dependent atmospheric extinction. From the standpoint of type Ia supernova cosmology, we interested in ensuring that our photometric zeropoints are consistent across the {\it ugrizy} bands, and that we properly account for any color perturbations due to atmospheric transmission variations. This motivates our focus on wavelength-dependent extinction differences {\it} across the four passbands.  

If we knew, at the mmag level, the product of the photon spectrum of the source (at the top of the atmosphere), and the throughput function of the telescope and instrument, we could extract the attenuation spectrum of the atmosphere from a single observation. We currently don't know the source spectrum or the instrument throughput at the requisite level. We therefore use the time-honored technique of using the airmass-dependence of the instrumental magnitudes to extract the atmosphere-dependent component of the measured magnitudes.  

For each of our four bands we fit a linear curve to the combined observations of that band across our airmass range:
\begin{equation}
    m_i = m_{i,0} + k_{i,\text{lin}}\times \chi.
\end{equation}
Here $m_i$ is the instrumental magnitude measured, $m_{i,0}$ is the top of the atmosphere instrumental magnitude for the specific star, while $k_{i,\text{lin}}$ is the linear extinction in that band and $\chi$ the airmass. From this we obtain an estimate of the top of the atmosphere instrumental magnitude of the star in each of the 4 bands. A caveat here is that we in fact split the data of the night into 3, because of the design of our survey, we measured the Quadband data on 3 different amplifiers of the camera, so we do a fit for each amplifier. See Table \ref{tab:TOAmag} for details. 
Using this fitted value(s), we can for any given image of the same star obtain an estimate of the extinction for the target star:
\begin{equation}
    k_i = \frac{m_i - m_{i,0}}{\chi}.
\end{equation}

\begin{table}[]
    \centering
    \caption{Top of the atmosphere magnitudes $(m_{i,0})$ estimated for HD2811 from observations on UT 20241016. See Figure \ref{fig:instrumentalmags}.}
    \label{tab:TOAmag}
    \begin{tabular}{ccccc}
    \hline
        amplifier & TOA band 1 & TOA band 2 & TOA band 3 & TOA band 4 \\ 
\hline
2 & $-14.4819\pm 0.000776$  & $-15.4839 \pm 0.00085 $  & $-14.1618 \pm 0.000861 $  & $-14.8106 \pm 0.000888 $\\ 
3 & $-14.4905\pm 0.000742$  & $-15.4883 \pm 0.00057 $  & $-14.1673 \pm 0.000841 $  & $-14.8154 \pm 0.000623 $\\ 
4 & $-14.4972\pm 0.000777$  & $-15.494 \pm 0.000918 $  & $-14.1745 \pm 0.000726 $  & $-14.8218 \pm 0.000941 $\\ 
        \hline
    \end{tabular}
    
\end{table}

\begin{table}[]
    \centering
    \caption{The extinctions measured $(k_{i,\text{lin}})$ for HD2811 from observations on UT 202241016. See Figure \ref{fig:instrumentalmags}}
    \label{tab:Extincs}
    \begin{tabular}{ccccc}
amplifier & Extinction band 1 & Extinction band 2 & Extinction band 3 & Extinction band 4 \\ 
\hline
2 & $0.400698\pm 0.000585$  & $0.219289 \pm 0.000603 $  & $0.14558 \pm 0.000638 $  & $0.124898 \pm 0.000606 $\\ 
3 & $0.404778\pm 0.000546$  & $0.220075 \pm 0.0004 $  & $0.146466 \pm 0.000637 $  & $0.125554 \pm 0.000476 $\\ 
4 & $0.403726\pm 0.000567$  & $0.217824 \pm 0.000704 $  & $0.145779 \pm 0.00054 $  & $0.124302 \pm 0.000724 $\\ 
\hline
    \end{tabular}
    
\end{table}

\begin{figure}
    \centering
    \includegraphics[width=\linewidth]{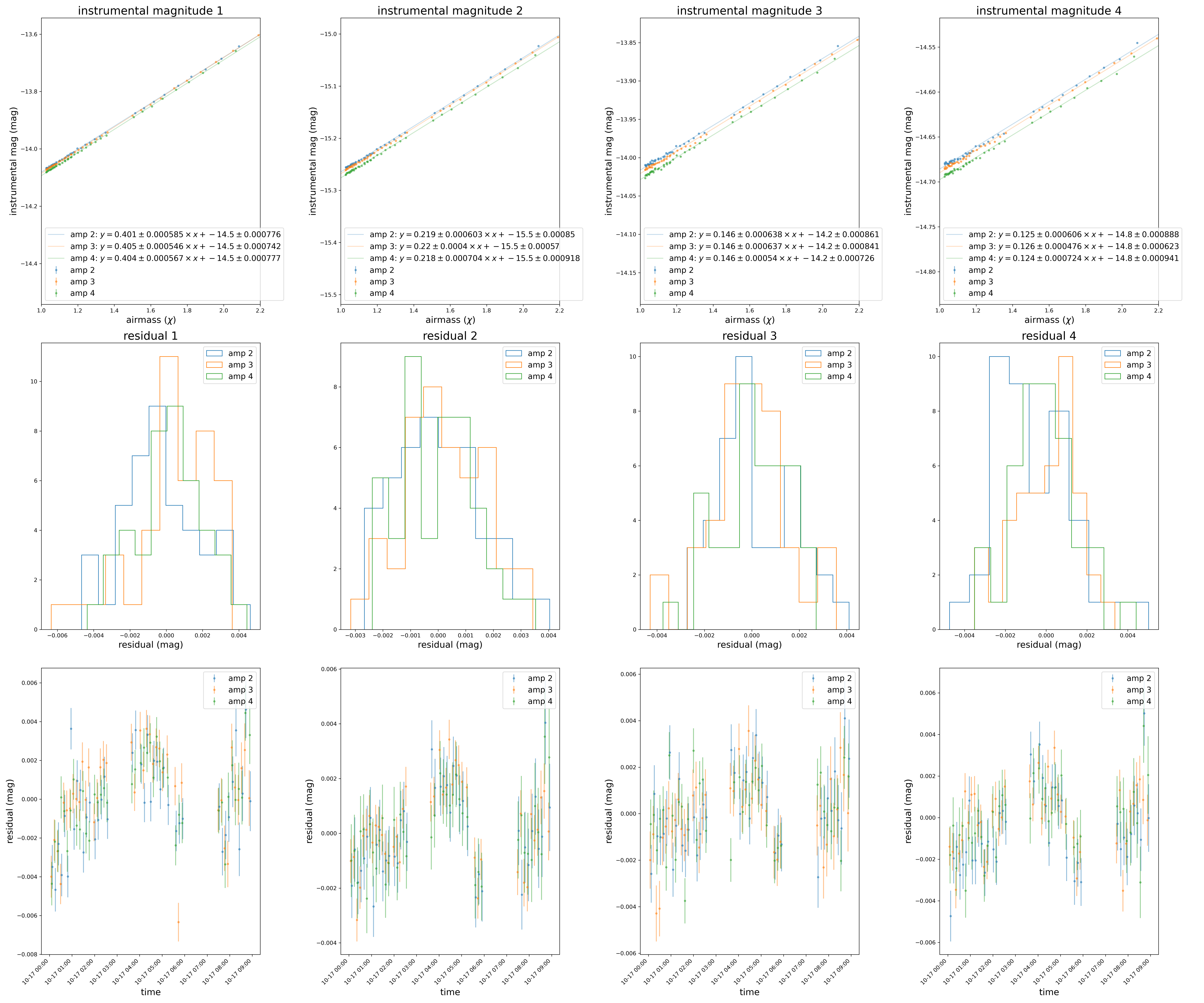}
    \caption{The fits for the four bands, including the residuals from the fits. For each band, we fit separately for each amplifier. The top row shows the fits and the measured magnitudes. The second row shows the histogram of the residuals in that band. The bottom row shows the residuals including the uncertainty as a function of time in the night.}
    \label{fig:instrumentalmags}
\end{figure}

\section{Analysis} \label{sec:analysis}

With the band-by-band extinctions found, we next analyze these to find the aerosol content. As mentioned earlier, there are several components that make up the atmospheric extinction in this wavelength regime:
\begin{equation}
    k_\text{Atmosphere} = k_\text{Aerosol} + k_\text{gray} + k_\text{Rayleigh} + k_\text{Ozone}.
\end{equation}
Here $k_\text{Atmosphere}$ is the entire extinction we found, $k_\text{Aerosol}$ is the aerosol component we want to find, $k_\text{gray}$ is the gray component caused by clouds, assumed to be independent of wavelengths, and $k_\text{Rayleigh}$ is the Rayleigh scattering term that is dependent on pressure, and is proportional to $\lambda^{-4}$, and finally $k_\text{Ozone}$ the Ozone extinction with a non-trivial wavelength dependence. The last two of these terms can be measured / predicted by sources other than the Rubin Observatory itself. The Rayleigh term is easily predicted from knowledge of the pressure, since its wavelength dependence is well known. The Ozone contribution can be determined through satellite remote sensing. 
To estimate these two components we rely on NASA's Planetary Spectrum Generator \citep{Villanueva_2018}, which can model the atmosphere through which we are observing. The Planetary Spectrum Generator obtains information on Ozone and pressure from the MERRA2 database \citep{MERRA2}. 
We invoke the Planetary Spectrum Generator (PSG) for each night, using the values at midnight local time as our "standard" for a given night. PSG delivers a Transmittance curve that includes both the Ozone and the Rayleigh terms. We then convert the transmittance curve to an extinction curve recalling that the transmittance is $T=e^{-\tau}$, with $\tau$ being the optical depth, which is directly proportional to the extinction:
\begin{equation}
    k_\text{PSG,correction} = -1.086\times\log\left(T_{PSG}\right).
\end{equation}
We estimate the deterministic extinction contribution for each of our four bands by taking the values at the effective center of the band. 

Subtracting that from our measured extinction should leave only the gray (wavelength-independent) contribution, and the Aerosol component. Equation (\ref{eq:optical_depth}) translated to extinctions gives us
\begin{equation}
    k_\text{Aerosol} = k_{Aerosol, \lambda_0}(t) \left(\frac{\lambda}{\lambda_0}\right)^{-\alpha(t)}.
    \label{eq:Angstrom}
\end{equation}
  We will use $\lambda_0= 500\text{nm}$. So from Equation (\ref{eq:Angstrom}) we see we have 2 parameters, a magnitude parameter  $k_{Aerosol, \lambda_0}(t)$ which is the value of extinction at $\lambda_0$, and an exponent $\alpha$ that determines the wavelength dependence of the aerosol attenuation. 

\citet{Eck_1999} suggested that the Angstrom law as written in Equation (\ref{eq:Angstrom}) was not enough and a second order term might be introduced by rewriting it in $\log$ space:
\begin{equation}
\label{eq:Angstrom2}
    \log\left(k_\text{Aerosol}\right) = \log\left(k_{Aerosol, \lambda_0}(t)\right) -\alpha_\text{linear}(t) \log\left(\frac{\lambda}{\lambda_0}\right) + \alpha_\text{quadratic} \left[\log\left(\frac{\lambda}{\lambda_0}\right)\right]^2.
\end{equation}

Finally, we used Equation \ref{eq:Angstrom} with a constant term $k_\text{gray}$ to allow for a cloud offset in all 4 bands. 

\section{Results}
\begin{figure}
    \centering
    \includegraphics[width=0.5\linewidth]{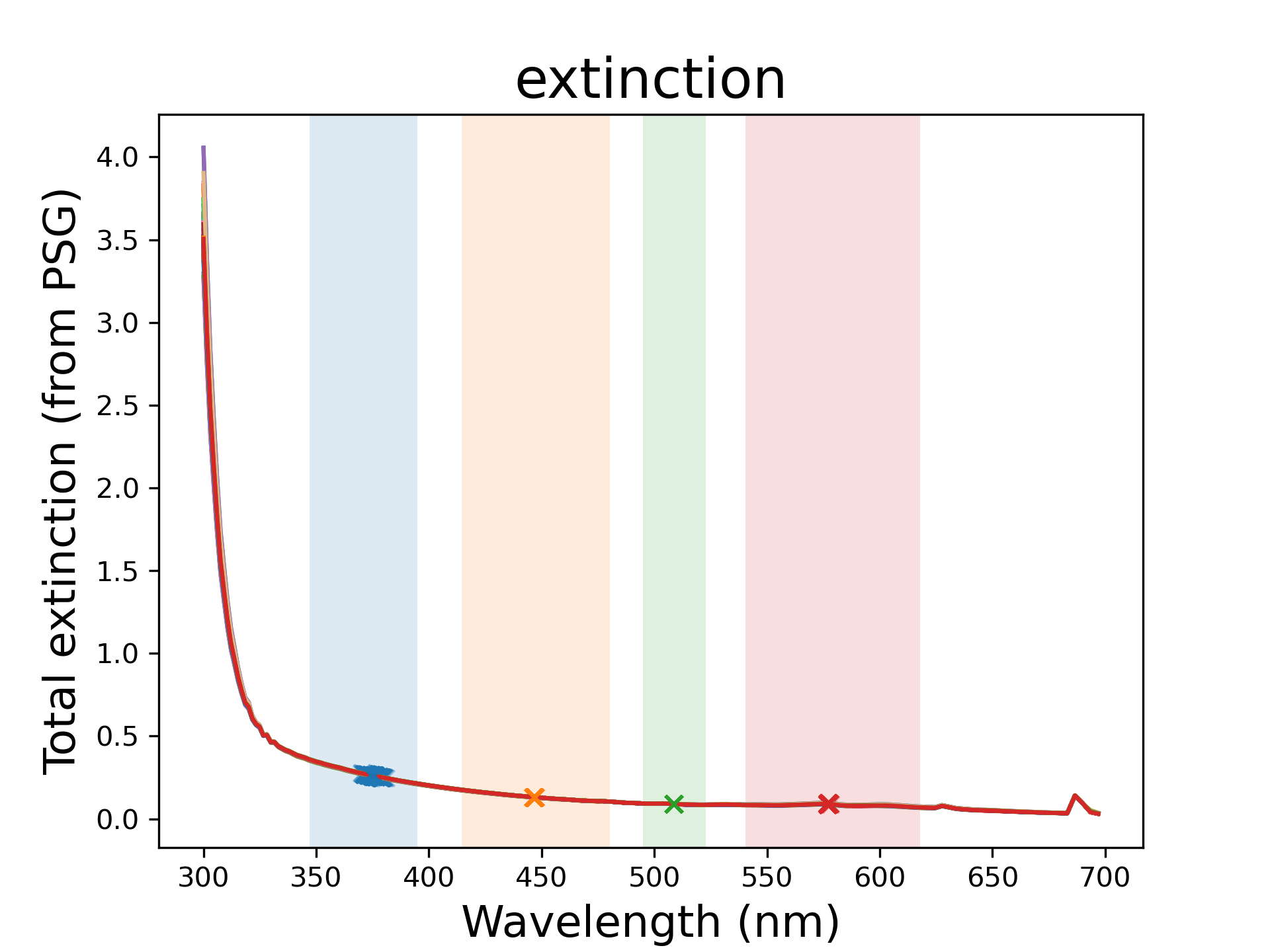}
    \caption{The combined Ozone and Rayleigh scattering extinction curves from PSG, for all nights. Also shown is the 4 bands indicated both by the entire band, and by the crosses on the line indicating the value at our best estimate of the effective center of the band.}
    \label{fig:PSG_extinction}
\end{figure}

For each night we observed HD2811, we used PSG to estimate an extinction curve for the Ozone and Rayleigh terms, as shown in Figure (\ref{fig:PSG_extinction}). 

\begin{figure}
    \centering
    \includegraphics[width=1\linewidth]{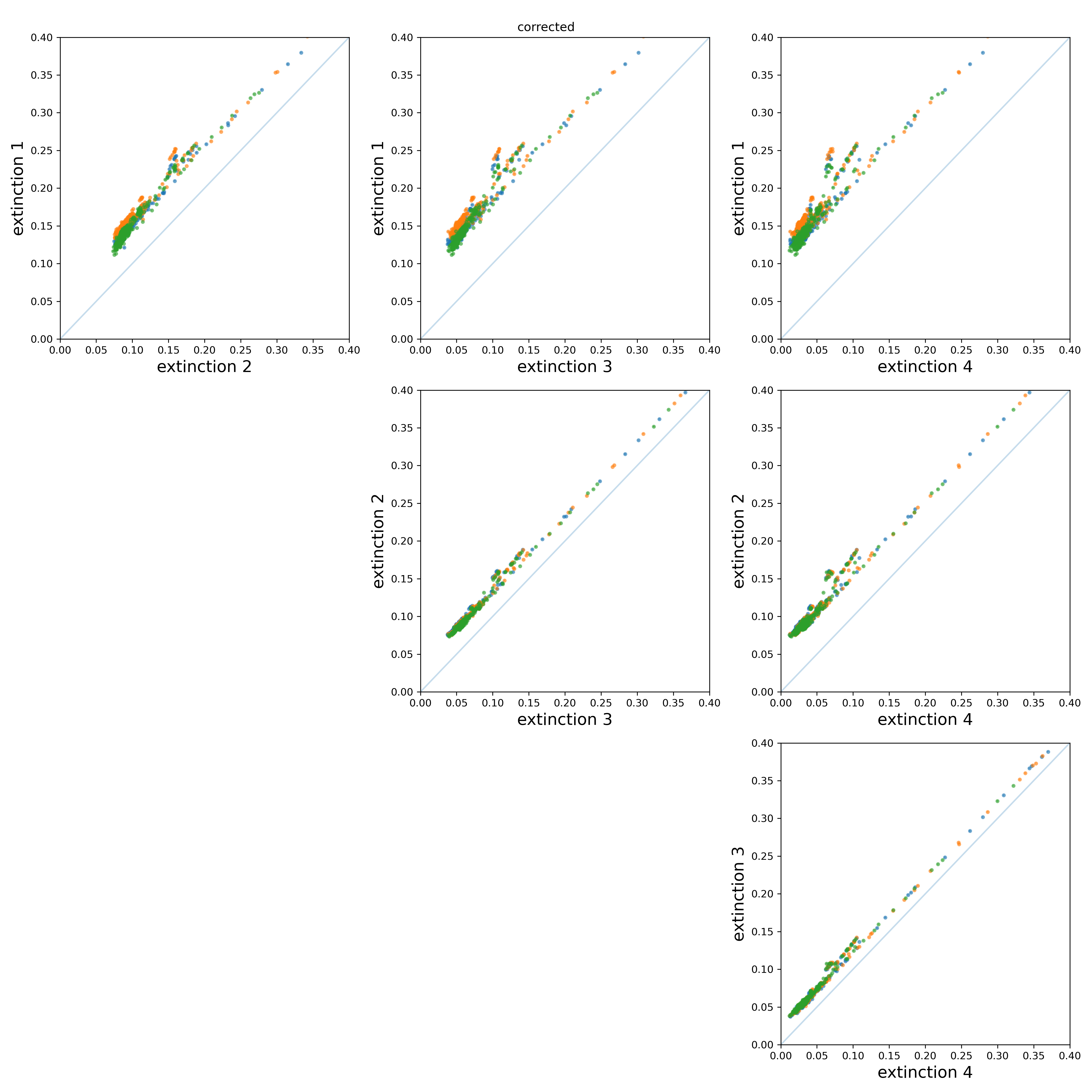}
    \caption{The excess extinction in each of the 4 bands plotted versus the excess extinction in the other bands. The colors indicate which of the 3 amplifiers that the spectrum landed on, which to us indicate that we might still have some issues with amplifiers we need to correct. An extinction along the diagonal line is what we would expect if there only was a cloud or gray component left in the excess extinction. While the offset and the indications of departures from the diagonal indicates wavelength dependent behavior. The fact that we can see more than one grouping of deviations could suggest more than one type of Aerosols, however we remind everyone that these are combined measurements conducted over nearly a year.}
    \label{fig:extincvsextinc}
\end{figure}

The excess extinction, beyond Rayleigh and Ozone, can be visualized different ways. We found one effective method to be comparing the extinction in the different bands, this is illustrated in Figure (\ref{fig:extincvsextinc}). Here we can see the extinction in the different bands follows lines mostly parallel with the diagonal, which is what would be expected if the excess extinction we are seeing here is purely caused by a gray term. The excursions from the 45-degree line arises from wavelength-dependent excess extinction, which we attribute to aerosols.

We fit various extinction models to the excess extinction in each image. 



\begin{figure}
    \centering
    \includegraphics[width=\linewidth]{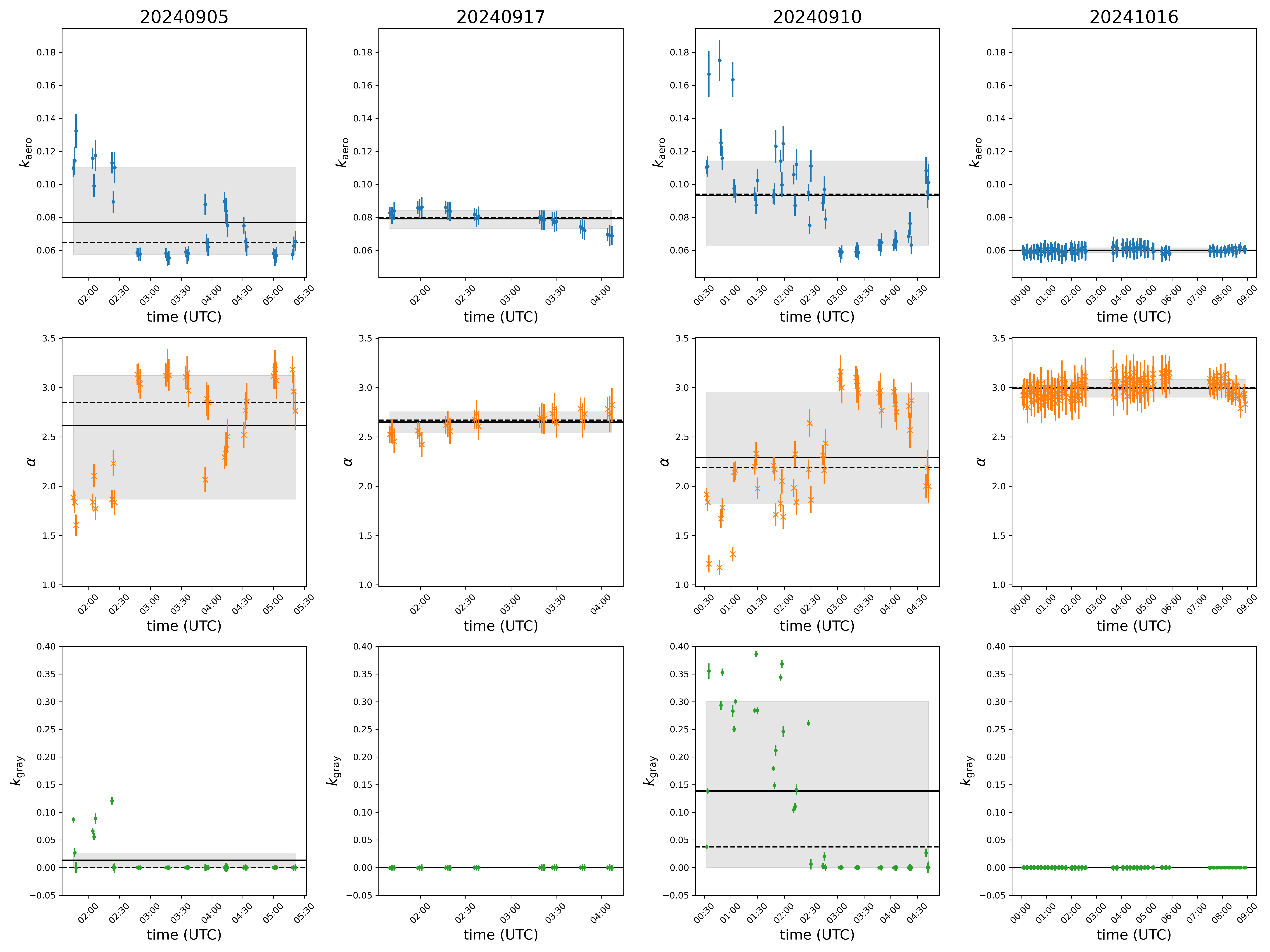}
    \caption{The fit parameters per image for the Angstrom model (Equation (\ref{eq:Angstrom})) plus a gray offset. We here picked out 4 examples of nights with both changing Aerosol content and possible Gray contamination. The top row shows the amplitude of the Aerosol extinction $k_\text{Aerosol}(\lambda_0=500\text{nm})$ across each of the nights. Second row shows the $\alpha$ parameter as a function of time in the night. The third row shows the gray offset $k_\text{gray}(\lambda_0=500\text{nm})$ again as a function of time. Each night we denoted the time in UTC. Also shown in each plot is the mean value (solid black), median value of the night (dashed line), and the 1 sigma range (gray area). The 4 nights chosen correspond to images number: 20240905 686-718, 20240917 818-841, 20240910 740-784, 20241016 1118-1243, for a figure of the complete sample see Figure \ref{fig:extincnight}. From this we see that there is night to night and even within night variations in parameters, but some nights are very stable.}
    \label{fig:extinctionfit_example}
\end{figure}

\begin{figure}
    \centering
    \includegraphics[width=1\linewidth]{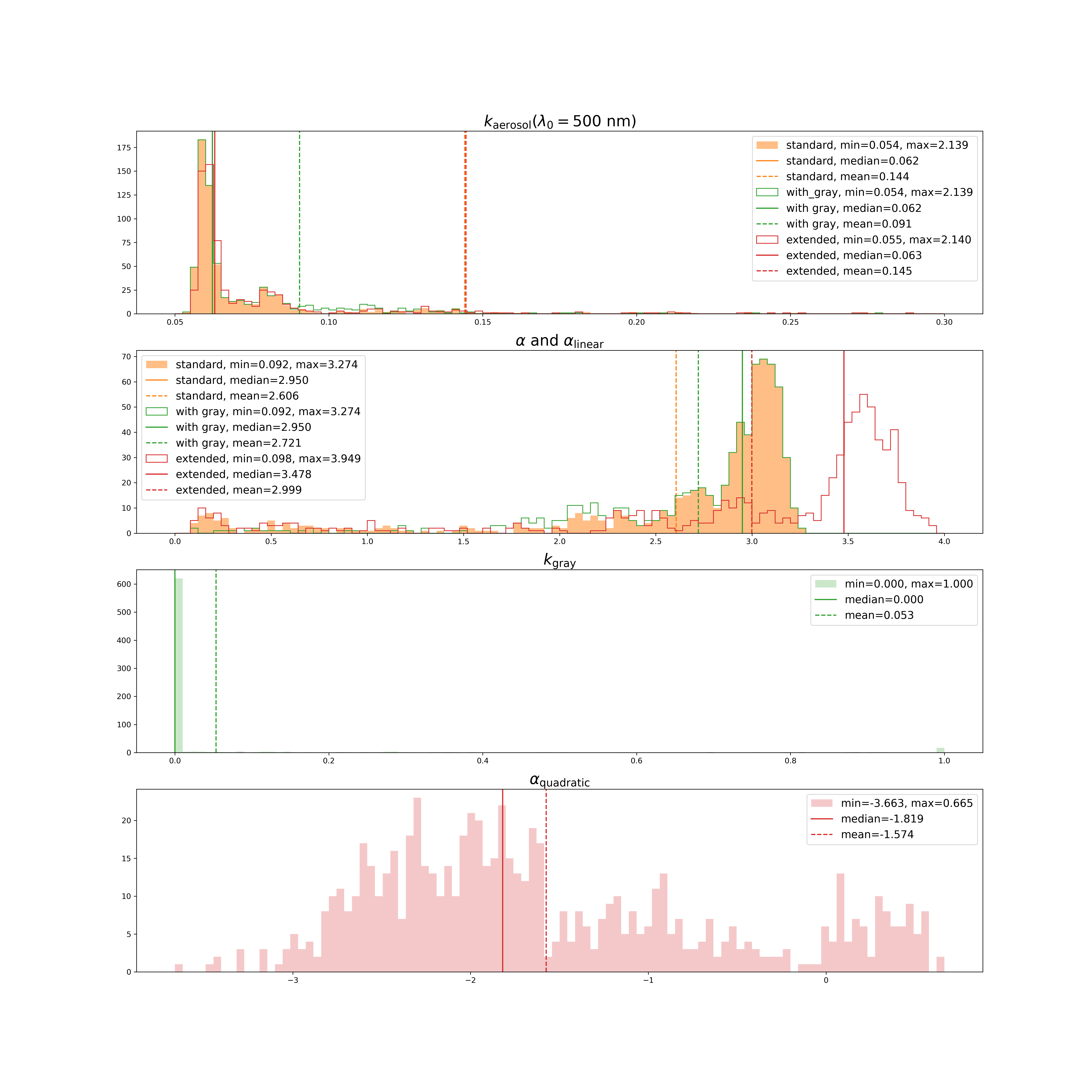}
    \caption{The histogram of the different parameters, In the top histogram, the amplitude $k_\text{Aerosol}(\lambda_0=500\text{nm})$, is plotted for all 3 fits, along with the mean and median of the distributions. Second graph shows the $\alpha$ parameter, for the Angstrom fits, both with (green) and without Gray terms (orange), and the linear $\alpha_\text{linear}$ from Equation (\ref{eq:Angstrom2}), again both the means and medians are also plotted. The 3rd plot shows the distribution of the $k_\text{gray}$ term, which can be seen to wary from 0 to 1. Finally the bottom plot shows the distribution of the $\alpha_\text{quadratic}$ term from Equation (\ref{eq:Angstrom2}).}
    \label{fig:histparam}
\end{figure}

To illustrate the data, we have plotted it in two different ways. Fit parameters as a function of time of night (in UTC) are shown in Figure (\ref{fig:extinctionfit_example}) for a few select nights (We only show details for the fit including a gray offset). We specifically illustrated in Figure (\ref{fig:extinctionfit_example}) nights where we saw variation but looking at the full picture in Figure (\ref{fig:extincnight}) we see that most nights are fairly stabile. 
Figure (\ref{fig:histparam}) shows a histogram of the of the fit parameters. 
We see that for the Angstrom model, that the distribution doesn't change much with inclusion of the $k_\text{gray}$ term. This can be seen in that the medians for both the amplitude $(k_\text{aerosol}(\lambda_0=500\text{nm}))$ and $\alpha$ are the same. 
However when looking at the mean values we notice that there does seem to be a shift caused by the inclusion of the gray term. The mean amplitude goes from 0.144 to 0.091 with the inclusion of a gray parameter. While the mean $\alpha$ parameter goes from 2.606 to 2.721 from the inclusion of the gray offset. 
When then comparing to the Quadratic model defined in Equation \ref{eq:Angstrom2} we see that the amplitude parameter is very similar in distribution, both with respect to mean and median, to that of the Angstrom model without the gray term. Meanwhile the $\alpha_\text{linear}$ is generally higher of value than either of the $\alpha$'s found with the regular Angstrom model. The latter can possibly be explained when we notice that the Quadratic term $(\alpha_\text{quadratic})$ tends towards the negative values, with both median and mean values being negative (-1.819 and -1.574 respectively), see the lowest graph in Figure \ref{fig:histparam}.
Finally we can look at the gray term $k_\text{gray}$ illustrated in Figure \ref{fig:histparam}'s 3rd plot, we see a median that is 0, while the mean is 0.053. 



\section{Conclusions} \label{sec:conclusion}

A shown in Table \ref{tab:Extincs}, the slitless dispersed quadband imager on Rubin's Auxiliary Telescope can determine atmospheric attenuation with a typical precision of 0.8 millimag/airmass. 

We seek to associate an atmospheric transmission curve with each Rubin observation. This entails obtaining summary parameters such as barometric pressure, Ozone content, precipitable water vapor, and aerosol coefficients $k$ and $\alpha$. These will inform a high-spectral-resolution model such as PSG to produce an atmospheric transmission spectrum that can be used in conjunction with the instrumental counterpart to perform forward-modeling spectrophotometry. 

An analysis of quadband dispersed imaging of HD2811 shows evidence of variation in wavelength-dependence of extinction. At the mmag level we will need to take account of color-airmass effects, and the precision we have achieved will enable this approach. 

Although we have ascribed the difference between the measured extinction and the calculated (Rayleigh + Ozone) contributions in terms of aerosols, we're essentially parameterizing any discrepancy between calculated and measured airmass-dependent continuum scattering, from any source. This polynomial parameterization will be included in a compact description of the atmosphere above the Rubin Observatory. 

\begin{acknowledgments}
EMP, CWS, and EU are grateful to both Harvard University and the US Department of Energy for their support of this work, the latter through the Cosmic Frontier grant DE-SC0007881. We thank the Rubin Observatory staff for their tireless efforts in acquiring Aux Tel data. 
Our colleagues Craig Lage, Fritz Muller, Tiago Ribeiro, and Mario Rivera were essential to making Aux Tel operational. 
The idea of the spatial mask was suggested by Marc Moniez and Martin Rodriguez Monroy, and we are grateful to them and also J\'er\'emy Neveu and Sylvie Dagoret-Campagne for their collaboration and partnership. This research has made use of the SIMBAD database,
operated at CDS, Strasbourg, France. We would also like to thank Geronimo Villanueva and his team at NASA for their help using the Planetary Spectrum Generator. 
\end{acknowledgments}

%

\vspace{5mm}
\facilities{Vera Rubin Auxtel}


\software{astropy\citep{astropy:2022, astropy:2018,astropy:2013}, PSG\citep{Villanueva_2018}, LSST Science Pipelines\citep{Rubin_pipelines}, 
          numpy\citep{harris2020array}, pandas\citep{reback2020pandas}, scipy\citep{2020SciPy-NMeth}, MODTRAN\citep{modtran6}, 
          }




\bibliography{QuadNotch}{}

\begin{thebibliography}{}
\expandafter\ifx\csname natexlab\endcsname\relax\def\natexlab#1{#1}\fi
\providecommand{\url}[1]{\href{#1}{#1}}
\providecommand{\dodoi}[1]{doi:~\href{http://doi.org/#1}{\nolinkurl{#1}}}
\providecommand{\doeprint}[1]{\href{http://ascl.net/#1}{\nolinkurl{http://ascl.net/#1}}}
\providecommand{\doarXiv}[1]{\href{https://arxiv.org/abs/#1}{\nolinkurl{https://arxiv.org/abs/#1}}}

\bibitem[{ {Astropy Collaboration} {et~al.}(2013){Astropy Collaboration}, {Robitaille}, {Tollerud}, {Greenfield}, {Droettboom}, {Bray}, {Aldcroft}, {Davis}, {Ginsburg}, {Price-Whelan}, {Kerzendorf}, {Conley}, {Crighton}, {Barbary}, {Muna}, {Ferguson}, {Grollier}, {Parikh}, {Nair}, {Unther}, {Deil}, {Woillez}, {Conseil}, {Kramer}, {Turner}, {Singer}, {Fox}, {Weaver}, {Zabalza}, {Edwards}, {Azalee Bostroem}, {Burke}, {Casey}, {Crawford}, {Dencheva}, {Ely}, {Jenness}, {Labrie}, {Lim}, {Pierfederici}, {Pontzen}, {Ptak}, {Refsdal}, {Servillat}, \& {Streicher}}]{astropy:2013}
{Astropy Collaboration}, {Robitaille}, T.~P., {Tollerud}, E.~J., {et~al.} 2013, \bibinfo{title}{{Astropy: A community Python package for astronomy},} \aap, 558, A33, \dodoi{10.1051/0004-6361/201322068}

\bibitem[{ {Astropy Collaboration} {et~al.}(2018){Astropy Collaboration}, {Price-Whelan}, {Sip{\H{o}}cz}, {G{\"u}nther}, {Lim}, {Crawford}, {Conseil}, {Shupe}, {Craig}, {Dencheva}, {Ginsburg}, {Vand erPlas}, {Bradley}, {P{\'e}rez-Su{\'a}rez}, {de Val-Borro}, {Aldcroft}, {Cruz}, {Robitaille}, {Tollerud}, {Ardelean}, {Babej}, {Bach}, {Bachetti}, {Bakanov}, {Bamford}, {Barentsen}, {Barmby}, {Baumbach}, {Berry}, {Biscani}, {Boquien}, {Bostroem}, {Bouma}, {Brammer}, {Bray}, {Breytenbach}, {Buddelmeijer}, {Burke}, {Calderone}, {Cano Rodr{\'\i}guez}, {Cara}, {Cardoso}, {Cheedella}, {Copin}, {Corrales}, {Crichton}, {D'Avella}, {Deil}, {Depagne}, {Dietrich}, {Donath}, {Droettboom}, {Earl}, {Erben}, {Fabbro}, {Ferreira}, {Finethy}, {Fox}, {Garrison}, {Gibbons}, {Goldstein}, {Gommers}, {Greco}, {Greenfield}, {Groener}, {Grollier}, {Hagen}, {Hirst}, {Homeier}, {Horton}, {Hosseinzadeh}, {Hu}, {Hunkeler}, {Ivezi{\'c}}, {Jain}, {Jenness}, {Kanarek}, {Kendrew}, {Kern}, {Kerzendorf}, {Khvalko}, {King}, {Kirkby}, {Kulkarni},
  {Kumar}, {Lee}, {Lenz}, {Littlefair}, {Ma}, {Macleod}, {Mastropietro}, {McCully}, {Montagnac}, {Morris}, {Mueller}, {Mumford}, {Muna}, {Murphy}, {Nelson}, {Nguyen}, {Ninan}, {N{\"o}the}, {Ogaz}, {Oh}, {Parejko}, {Parley}, {Pascual}, {Patil}, {Patil}, {Plunkett}, {Prochaska}, {Rastogi}, {Reddy Janga}, {Sabater}, {Sakurikar}, {Seifert}, {Sherbert}, {Sherwood-Taylor}, {Shih}, {Sick}, {Silbiger}, {Singanamalla}, {Singer}, {Sladen}, {Sooley}, {Sornarajah}, {Streicher}, {Teuben}, {Thomas}, {Tremblay}, {Turner}, {Terr{\'o}n}, {van Kerkwijk}, {de la Vega}, {Watkins}, {Weaver}, {Whitmore}, {Woillez}, {Zabalza}, \& {Astropy Contributors}}]{astropy:2018}
{Astropy Collaboration}, {Price-Whelan}, A.~M., {Sip{\H{o}}cz}, B.~M., {et~al.} 2018, \bibinfo{title}{{The Astropy Project: Building an Open-science Project and Status of the v2.0 Core Package},} \aj, 156, 123, \dodoi{10.3847/1538-3881/aabc4f}

\bibitem[{ {Astropy Collaboration} {et~al.}(2022){Astropy Collaboration}, {Price-Whelan}, {Lim}, {Earl}, {Starkman}, {Bradley}, {Shupe}, {Patil}, {Corrales}, {Brasseur}, {N{"o}the}, {Donath}, {Tollerud}, {Morris}, {Ginsburg}, {Vaher}, {Weaver}, {Tocknell}, {Jamieson}, {van Kerkwijk}, {Robitaille}, {Merry}, {Bachetti}, {G{"u}nther}, {Aldcroft}, {Alvarado-Montes}, {Archibald}, {B{'o}di}, {Bapat}, {Barentsen}, {Baz{'a}n}, {Biswas}, {Boquien}, {Burke}, {Cara}, {Cara}, {Conroy}, {Conseil}, {Craig}, {Cross}, {Cruz}, {D'Eugenio}, {Dencheva}, {Devillepoix}, {Dietrich}, {Eigenbrot}, {Erben}, {Ferreira}, {Foreman-Mackey}, {Fox}, {Freij}, {Garg}, {Geda}, {Glattly}, {Gondhalekar}, {Gordon}, {Grant}, {Greenfield}, {Groener}, {Guest}, {Gurovich}, {Handberg}, {Hart}, {Hatfield-Dodds}, {Homeier}, {Hosseinzadeh}, {Jenness}, {Jones}, {Joseph}, {Kalmbach}, {Karamehmetoglu}, {Ka{l}uszy{'n}ski}, {Kelley}, {Kern}, {Kerzendorf}, {Koch}, {Kulumani}, {Lee}, {Ly}, {Ma}, {MacBride}, {Maljaars}, {Muna}, {Murphy}, {Norman}, {O'Steen},
  {Oman}, {Pacifici}, {Pascual}, {Pascual-Granado}, {Patil}, {Perren}, {Pickering}, {Rastogi}, {Roulston}, {Ryan}, {Rykoff}, {Sabater}, {Sakurikar}, {Salgado}, {Sanghi}, {Saunders}, {Savchenko}, {Schwardt}, {Seifert-Eckert}, {Shih}, {Jain}, {Shukla}, {Sick}, {Simpson}, {Singanamalla}, {Singer}, {Singhal}, {Sinha}, {Sip{H{o}}cz}, {Spitler}, {Stansby}, {Streicher}, {{{S}}umak}, {Swinbank}, {Taranu}, {Tewary}, {Tremblay}, {Val-Borro}, {Van Kooten}, {Vasovi{'c}}, {Verma}, {de Miranda Cardoso}, {Williams}, {Wilson}, {Winkel}, {Wood-Vasey}, {Xue}, {Yoachim}, {Zhang}, {Zonca}, \& {Astropy Project Contributors}}]{astropy:2022}
{Astropy Collaboration}, {Price-Whelan}, A.~M., {Lim}, P.~L., {et~al.} 2022, \bibinfo{title}{{The Astropy Project: Sustaining and Growing a Community-oriented Open-source Project and the Latest Major Release (v5.0) of the Core Package},} \apj, 935, 167, \dodoi{10.3847/1538-4357/ac7c74}

\bibitem[{A. Berk {et~al.}(2014)Berk, Conforti, Kennett, Perkins, Hawes, \& van~den Bosch}]{modtran6}
Berk, A., Conforti, P., Kennett, R., {et~al.} 2014, \bibinfo{title}{{MODTRAN6: a major upgrade of the MODTRAN radiative transfer code},} in Algorithms and Technologies for Multispectral, Hyperspectral, and Ultraspectral Imagery XX, ed. M.~Velez-Reyes \& F.~A. Kruse, Vol. 9088, International Society for Optics and Photonics (SPIE), 90880H, \dodoi{10.1117/12.2050433}

\bibitem[{G. Bernstein {et~al.}(2017)Bernstein, Abbott, Desai, Gruen, Gruendl, Johnson, Lin, Menanteau, Morganson, Neilsen, {et~al.}}]{bernstein2017instrumental}
Bernstein, G., Abbott, T., Desai, S., {et~al.} 2017, \bibinfo{title}{Instrumental response model and detrending for the Dark Energy Camera,} Publications of the Astronomical Society of the Pacific, 129, 114502

\bibitem[{C.~H. Blake \& M.~M. Shaw(2011)Blake \& Shaw}]{blake2011measuring}
Blake, C.~H., \& Shaw, M.~M. 2011, \bibinfo{title}{Measuring NIR Atmospheric Extinction Using a Global Positioning System Receiver,} Publications of the Astronomical Society of the Pacific, 123, 1302

\bibitem[{R.~C. {Bohlin} {et~al.}(2020){Bohlin}, {Hubeny}, \& {Rauch}}]{CalSpec2020}
{Bohlin}, R.~C., {Hubeny}, I., \& {Rauch}, T. 2020, \bibinfo{title}{{New Grids of Pure-hydrogen White Dwarf NLTE Model Atmospheres and the HST/STIS Flux Calibration},} \aj, 160, 21, \dodoi{10.3847/1538-3881/ab94b4}

\bibitem[{D. Burke {et~al.}(2017)Burke, Rykoff, Allam, Annis, Bechtol, Bernstein, Drlica-Wagner, Finley, Gruendl, James, {et~al.}}]{burke2017forward}
Burke, D., Rykoff, E., Allam, S., {et~al.} 2017, \bibinfo{title}{Forward global photometric calibration of the dark energy survey,} The Astronomical Journal, 155, 41

\bibitem[{D.~L. Burke {et~al.}(2010)Burke, Axelrod, Blondin, Claver, Ivezi{\'c}, Jones, Saha, Smith, Smith, \& Stubbs}]{burke2010precision}
Burke, D.~L., Axelrod, T., Blondin, S., {et~al.} 2010, \bibinfo{title}{Precision determination of atmospheric extinction at optical and near-infrared wavelengths,} The Astrophysical Journal, 720, 811

\bibitem[{D.~L. Burke {et~al.}(2013)Burke, Saha, Claver, Axelrod, Claver, DePoy, Ivezi{\'c}, Jones, Smith, \& Stubbs}]{burke2013all}
Burke, D.~L., Saha, A., Claver, J., {et~al.} 2013, \bibinfo{title}{All-weather calibration of wide-field optical and NIR surveys,} The Astronomical Journal, 147, 19

\bibitem[{C. Buton {et~al.}(2013)Buton, Copin, Aldering, Antilogus, Aragon, Bailey, Baltay, Bongard, Canto, Cellier-Holzem, {et~al.}}]{buton2013atmospheric}
Buton, C., Copin, Y., Aldering, G., {et~al.} 2013, \bibinfo{title}{Atmospheric extinction properties above Mauna Kea from the Nearby SuperNova Factory spectro-photometric data set,} Astronomy \& Astrophysics, 549, A8

\bibitem[{M.~W. Coughlin {et~al.}(2018)Coughlin, Deustua, Guyonnet, Mondrik, Rice, Stubbs, \& Woodward}]{coughlin2018testing}
Coughlin, M.~W., Deustua, S., Guyonnet, A., {et~al.} 2018, \bibinfo{title}{Testing of the LSST's photometric calibration strategy at the CTIO 0.9 meter telescope,} in Observatory Operations: Strategies, Processes, and Systems VII, Vol. 10704, SPIE, 753--765

\bibitem[{M. Dawsey {et~al.}(2006)Dawsey, Gimmestad, Roberts, McGraw, Zimmer, \& Fitch}]{dawsey2006lidar}
Dawsey, M., Gimmestad, G., Roberts, D., {et~al.} 2006, \bibinfo{title}{LIDAR for measuring atmospheric extinction,} in Observatory Operations: Strategies, Processes, and Systems, Vol. 6270, SPIE, 453--462

\bibitem[{T.~F. Eck {et~al.}(1999)Eck, Holben, Reid, Dubovik, Smirnov, O'Neill, Slutsker, \& Kinne}]{Eck_1999}
Eck, T.~F., Holben, B.~N., Reid, J.~S., {et~al.} 1999, \bibinfo{title}{Wavelength dependence of the optical depth of biomass burning, urban, and desert dust aerosols,} Journal of Geophysical Research: Atmospheres, 104, 31333, \dodoi{https://doi.org/10.1029/1999JD900923}

\bibitem[{K. Gilmore {et~al.}(2008)Gilmore, Kahn, Nordby, O'Connor, Oliver, Radeka, Schalk, Schindler, \& Van~Berg}]{gilmore2008lsst}
Gilmore, K., Kahn, S., Nordby, M., {et~al.} 2008, \bibinfo{title}{The LSST camera overview: design and performance,} in Ground-based and Airborne Instrumentation for Astronomy II, Vol. 7014, SPIE, 167--177

\bibitem[{A. Guyonnet {et~al.}(2019)Guyonnet, Dagoret-Campagne, \& Mondrik}]{guyonnet2019local}
Guyonnet, A., Dagoret-Campagne, S., \& Mondrik, N. 2019, \bibinfo{title}{Local monitoring of atmospheric transparency from the NASA MERRA-2 global assimilation system,} Journal of Astronomical Instrumentation, 8, 1950013

\bibitem[{C.~R. Harris {et~al.}(2020)Harris, Millman, van~der Walt, Gommers, Virtanen, Cournapeau, Wieser, Taylor, Berg, Smith, Kern, Picus, Hoyer, van Kerkwijk, Brett, Haldane, del R{\'{i}}o, Wiebe, Peterson, G{\'{e}}rard-Marchant, Sheppard, Reddy, Weckesser, Abbasi, Gohlke, \& Oliphant}]{harris2020array}
Harris, C.~R., Millman, K.~J., van~der Walt, S.~J., {et~al.} 2020, \bibinfo{title}{Array programming with {NumPy},} Nature, 585, 357, \dodoi{10.1038/s41586-020-2649-2}

\bibitem[{B.~N. Holben {et~al.}(1998)Holben, Eck, Slutsker, Tanr{\'e}, Buis, Setzer, Vermote, Reagan, Kaufman, Nakajima, {et~al.}}]{holben1998aeronet}
Holben, B.~N., Eck, T.~F., Slutsker, I.~a., {et~al.} 1998, \bibinfo{title}{AERONET—A federated instrument network and data archive for aerosol characterization,} Remote sensing of environment, 66, 1

\bibitem[{P. Ingraham {et~al.}(2022)Ingraham, Fagrelius, Stubbs, Lupton, Liang, Neill, Muller, Claver, Sebag, Thomas, {et~al.}}]{ingraham2022vera}
Ingraham, P., Fagrelius, P., Stubbs, C.~W., {et~al.} 2022, \bibinfo{title}{The Vera C. Rubin Observatory 8.4 m telescope calibration system status,} in Ground-based and Airborne Telescopes IX, Vol. 12182, SPIE, 272--285

\bibitem[{{\v{Z}}. Ivezi{\'c} {et~al.}(2019)Ivezi{\'c}, Kahn, Tyson, Abel, Acosta, Allsman, Alonso, AlSayyad, Anderson, Andrew, {et~al.}}]{ivezic2019lsst}
Ivezi{\'c}, {\v{Z}}., Kahn, S.~M., Tyson, J.~A., {et~al.} 2019, \bibinfo{title}{LSST: from science drivers to reference design and anticipated data products,} The Astrophysical Journal, 873, 111

\bibitem[{M. Lesser \& D. Ouellette(2017)Lesser \& Ouellette}]{lesser2017results}
Lesser, M., \& Ouellette, D. 2017, \bibinfo{title}{Results from STA/ITL fully depleted CCDs for LSST,} Journal of Instrumentation, 12, C03080

\bibitem[{D. Li {et~al.}(2017)Li, Blake, Nidever, \& Halverson}]{li2017temporal}
Li, D., Blake, C.~H., Nidever, D., \& Halverson, S.~P. 2017, \bibinfo{title}{Temporal Variations of Telluric Water Vapor Absorption at Apache Point Observatory,} Publications of the Astronomical Society of the Pacific, 130, 014501

\bibitem[{T. Li {et~al.}(2014)Li, DePoy, Marshall, Nagasawa, Carona, \& Boada}]{li2014monitoring}
Li, T., DePoy, D., Marshall, J.~L., {et~al.} 2014, \bibinfo{title}{Monitoring the atmospheric throughput at Cerro Tololo Inter-American Observatory with aTmCam,} in Ground-based and Airborne Instrumentation for Astronomy V, Vol. 9147, SPIE, 2194--2205

\bibitem[{T. Li {et~al.}(2012)Li, DePoy, Kessler, Burke, Marshall, Wise, Rheault, Carona, Boada, Prochaska, {et~al.}}]{li2012atmcam}
Li, T., DePoy, D., Kessler, R., {et~al.} 2012, \bibinfo{title}{aTmcam: a simple atmospheric transmission monitoring camera for sub 1\% photometric precision,} in Ground-based and Airborne Instrumentation for Astronomy IV, Vol. 8446, SPIE, 922--934

\bibitem[{T. Li {et~al.}(2016)Li, DePoy, Marshall, Tucker, Kessler, Annis, Bernstein, Boada, Burke, Finley, {et~al.}}]{li2016assessment}
Li, T., DePoy, D., Marshall, J., {et~al.} 2016, \bibinfo{title}{Assessment of systematic chromatic errors that impact sub-1\% photometric precision in large-area sky surveys,} The Astronomical Journal, 151, 157

\bibitem[{K. Louedec(2015)Louedec}]{louedec2015atmospheric}
Louedec, K. 2015, \bibinfo{title}{Atmospheric effects in astroparticle physics experiments and the challenge of ever greater precision in measurements,} Astroparticle Physics, 60, 54

\bibitem[{J.~T. McGraw {et~al.}(2009)McGraw, Stubbs, Zimmer, Fraser, \& Vivekanandan}]{mcgraw2009measurement}
McGraw, J.~T., Stubbs, C.~W., Zimmer, P.~C., Fraser, G.~T., \& Vivekanandan, J. 2009, \bibinfo{title}{Measurement astrophysics (MAP) first steps: A new decade of ground-based photometric precision and accuracy,} Astro2010: The Astronomy and Astrophysics Decadal Survey, Technology Development Papers

\bibitem[{N.~P. Mondrik(2020)Mondrik}]{mondrik2020calibration}
Mondrik, N.~P. 2020, \bibinfo{title}{Calibration Hardware and Methodology for Large Photometric Surveys,} PhD thesis, Harvard University

\bibitem[{M. Moniez {et~al.}(2021)Moniez, Neveu, Dagoret-Campagne, Gentet, Le~Guillou, \& M}]{moniez2021transmission}
Moniez, M., Neveu, J., Dagoret-Campagne, S., {et~al.} 2021, \bibinfo{title}{A transmission hologram for slitless spectrophotometry on a convergent telescope beam. 1. Focus and resolution,} Monthly Notices of the Royal Astronomical Society, 506, 5589

\bibitem[{ NASA(2025)NASA}]{MERRA2}
NASA. 2025, Modern-Era Retrospective analysis for Research and Applications, Version 2, \url{https://gmao.gsfc.nasa.gov/gmao-products/merra-2/}

\bibitem[{J. Neveu {et~al.}(2023)Neveu, Br{\'e}maud, Antilogus, Barret, Bongard, Copin, Dagoret-Campagne, Juramy, Le-Guillou, Moniez, {et~al.}}]{neveu2023slitless}
Neveu, J., Br{\'e}maud, V., Antilogus, P., {et~al.} 2023, \bibinfo{title}{Slitless spectrophotometry with forward modelling: principles and application to atmospheric transmission measurement,} arXiv preprint arXiv:2307.04898

\bibitem[{T. pandas~development team(2020)pandas~development team}]{reback2020pandas}
pandas~development team, T. 2020, pandas-dev/pandas: Pandas, latest Zenodo, \dodoi{10.5281/zenodo.3509134}

\bibitem[{F. Patat {et~al.}(2011)Patat, Moehler, O'Brien, Pompei, Bensby, Carraro, de~Ugarte~Postigo, Fox, Gavignaud, James, {et~al.}}]{patat2011optical}
Patat, F., Moehler, S., O'Brien, K., {et~al.} 2011, \bibinfo{title}{Optical atmospheric extinction over Cerro Paranal,} Astronomy \& Astrophysics, 527, A91

\bibitem[{E. Peretz {et~al.}(2021)Peretz, Hamilton, Mather, Pabarcius, Hall, Michaels, Pritchett, Yu, Wizinowich, \& Golliher}]{peretz2021orcas}
Peretz, E., Hamilton, C., Mather, J.~C., {et~al.} 2021, \bibinfo{title}{ORCAS--orbiting configurable artificial star mission architecture,} in UV/Optical/IR Space Telescopes and Instruments: Innovative Technologies and Concepts X, Vol. 11819, SPIE, 26--40

\bibitem[{S. Perlmutter {et~al.}(1999)Perlmutter, Aldering, Goldhaber, Knop, Nugent, Castro, Deustua, Fabbro, Goobar, Groom, {et~al.}}]{perlmutter1999measurements}
Perlmutter, S., Aldering, G., Goldhaber, G., {et~al.} 1999, \bibinfo{title}{Measurements of $\Omega$ and $\Lambda$ from 42 high-redshift supernovae,} The Astrophysical Journal, 517, 565

\bibitem[{H.-G. Reimann \& S. Beyersdorfer(1992)Reimann \& Beyersdorfer}]{reimann1992atmospheric}
Reimann, H.-G., \& Beyersdorfer, S. 1992, \bibinfo{title}{Atmospheric extinction and meteorological conditions-A long time photometric study,} Astronomy and Astrophysics (ISSN 0004-6361), vol. 265, no. 1, p. 360-369., 265, 360

\bibitem[{A.~G. Riess {et~al.}(1998)Riess, Filippenko, Challis, Clocchiatti, Diercks, Garnavich, Gilliland, Hogan, Jha, Kirshner, {et~al.}}]{riess1998observational}
Riess, A.~G., Filippenko, A.~V., Challis, P., {et~al.} 1998, \bibinfo{title}{Observational evidence from supernovae for an accelerating universe and a cosmological constant,} The astronomical journal, 116, 1009

\bibitem[{C.~W. Stubbs \& Y.~J. Brown(2015)Stubbs \& Brown}]{stubbs2015precise}
Stubbs, C.~W., \& Brown, Y.~J. 2015, \bibinfo{title}{Precise astronomical flux calibration and its impact on studying the nature of the dark energy,} Modern Physics Letters A, 30, 1530030

\bibitem[{C.~W. Stubbs \& J.~L. Tonry(2006)Stubbs \& Tonry}]{stubbs2006toward}
Stubbs, C.~W., \& Tonry, J.~L. 2006, \bibinfo{title}{Toward 1\% photometry: End-to-end calibration of astronomical telescopes and detectors,} The Astrophysical Journal, 646, 1436

\bibitem[{C.~W. Stubbs {et~al.}(2015)Stubbs, Vaz, Fraser, Cramer, Lykke, \& Woodward}]{stubbs2015comparison}
Stubbs, C.~W., Vaz, A., Fraser, G.~T., {et~al.} 2015, \bibinfo{title}{Comparison of MODTRAN5 atmospheric extinction predictions with narrowband astronomical flux observations,} in Infrared Remote Sensing and Instrumentation XXIII, Vol. 9608, SPIE, 214--223

\bibitem[{C.~W. Stubbs {et~al.}(2007)Stubbs, High, George, DeRose, Blondin, Tonry, Chambers, Granett, Burke, \& Smith}]{stubbs2007toward}
Stubbs, C.~W., High, F.~W., George, M.~R., {et~al.} 2007, \bibinfo{title}{Toward more precise survey photometry for PanSTARRS and LSST: measuring directly the optical transmission spectrum of the atmosphere,} Publications of the Astronomical Society of the Pacific, 119, 1163

\bibitem[{D.~L. Tucker {et~al.}(2006)Tucker, Kent, Richmond, Annis, Smith, Allam, Rodgers, Stute, Adelman-McCarthy, Brinkmann, {et~al.}}]{tucker2006sloan}
Tucker, D.~L., Kent, S., Richmond, M., {et~al.} 2006, \bibinfo{title}{The sloan digital sky survey monitor telescope pipeline,} Astronomische Nachrichten: Astronomical Notes, 327, 821

\bibitem[{ {Vera C. Rubin Observatory Science Pipelines Developers}(2025){Vera C. Rubin Observatory Science Pipelines Developers}}]{Rubin_pipelines}
{Vera C. Rubin Observatory Science Pipelines Developers}. 2025, {The LSST Science Pipelines Software: Optical Survey Pipeline Reduction and Analysis Environment}, {Project Science Technical Note} PSTN-019, {Vera C. Rubin Observatory}, \dodoi{10.71929/rubin/2570545}

\bibitem[{G. Villanueva {et~al.}(2018)Villanueva, Smith, Protopapa, Faggi, \& Mandell}]{Villanueva_2018}
Villanueva, G., Smith, M., Protopapa, S., Faggi, S., \& Mandell, A. 2018, \bibinfo{title}{Planetary Spectrum Generator: An accurate online radiative transfer suite for atmospheres, comets, small bodies and exoplanets,} Journal of Quantitative Spectroscopy and Radiative Transfer, 217, 86–104, \dodoi{10.1016/j.jqsrt.2018.05.023}

\bibitem[{P. Virtanen {et~al.}(2020)Virtanen, Gommers, Oliphant, Haberland, Reddy, Cournapeau, Burovski, Peterson, Weckesser, Bright, {van der Walt}, Brett, Wilson, Millman, Mayorov, Nelson, Jones, Kern, Larson, Carey, Polat, Feng, Moore, {VanderPlas}, Laxalde, Perktold, Cimrman, Henriksen, Quintero, Harris, Archibald, Ribeiro, Pedregosa, {van Mulbregt}, \& {SciPy 1.0 Contributors}}]{2020SciPy-NMeth}
Virtanen, P., Gommers, R., Oliphant, T.~E., {et~al.} 2020, \bibinfo{title}{{{SciPy} 1.0: Fundamental Algorithms for Scientific Computing in Python},} Nature Methods, 17, 261, \dodoi{10.1038/s41592-019-0686-2}

\bibitem[{M. {Wenger} {et~al.}(2000){Wenger}, {Ochsenbein}, {Egret}, {Dubois}, {Bonnarel}, {Borde}, {Genova}, {Jasniewicz}, {Lalo{\"e}}, {Lesteven}, \& {Monier}}]{wenger2000simbad}
{Wenger}, M., {Ochsenbein}, F., {Egret}, D., {et~al.} 2000, \bibinfo{title}{{The SIMBAD astronomical database. The CDS reference database for astronomical objects},} \aaps, 143, 9, \dodoi{10.1051/aas:2000332}

\bibitem[{W. Wood-Vasey {et~al.}(2022)Wood-Vasey, Perrefort, \& Baker}]{wood2022gps}
Wood-Vasey, W., Perrefort, D., \& Baker, A.~D. 2022, \bibinfo{title}{GPS Measurements of Precipitable Water Vapor Can Improve Survey Calibration: A Demonstration from KPNO and the Mayall z-band Legacy Survey,} The Astronomical Journal, 163, 283

\bibitem[{A.~T. Young {et~al.}(1991)Young, Genet, Boyd, Borucki, Lockwood, Henry, Hall, Smith, Baliumas, Donahue, {et~al.}}]{young1991precise}
Young, A.~T., Genet, R.~M., Boyd, L.~J., {et~al.} 1991, \bibinfo{title}{Precise automatic differential stellar photometry,} Publications of the Astronomical Society of the Pacific, 103, 221

\bibitem[{P. Zimmer {et~al.}(2012)Zimmer, McGraw, Zirzow, Cramer, Lykke, \& Woodward~IV}]{zimmer2012new}
Zimmer, P., McGraw, J.~T., Zirzow, D.~C., {et~al.} 2012, \bibinfo{title}{New instruments to calibrate atmospheric transmission,} in Ground-based and Airborne Telescopes IV, Vol. 8444, SPIE, 573--584

\bibitem[{P.~C. Zimmer {et~al.}(2006)Zimmer, McGraw, Gimmestad, Roberts, Stewart, Dawsey, Fitch, Smith, Townsend, \& Black}]{zimmer2006ale}
Zimmer, P.~C., McGraw, J., Gimmestad, G., {et~al.} 2006, \bibinfo{title}{ALE: Astronomical LIDAR for Extinction,} in American Astronomical Society Meeting Abstracts, Vol. 209, 154--04

\end{thebibliography}
\bibliographystyle{aasjournalv7}

\appendix

\begin{figure}
    \centering
    \includegraphics[width=1\linewidth]{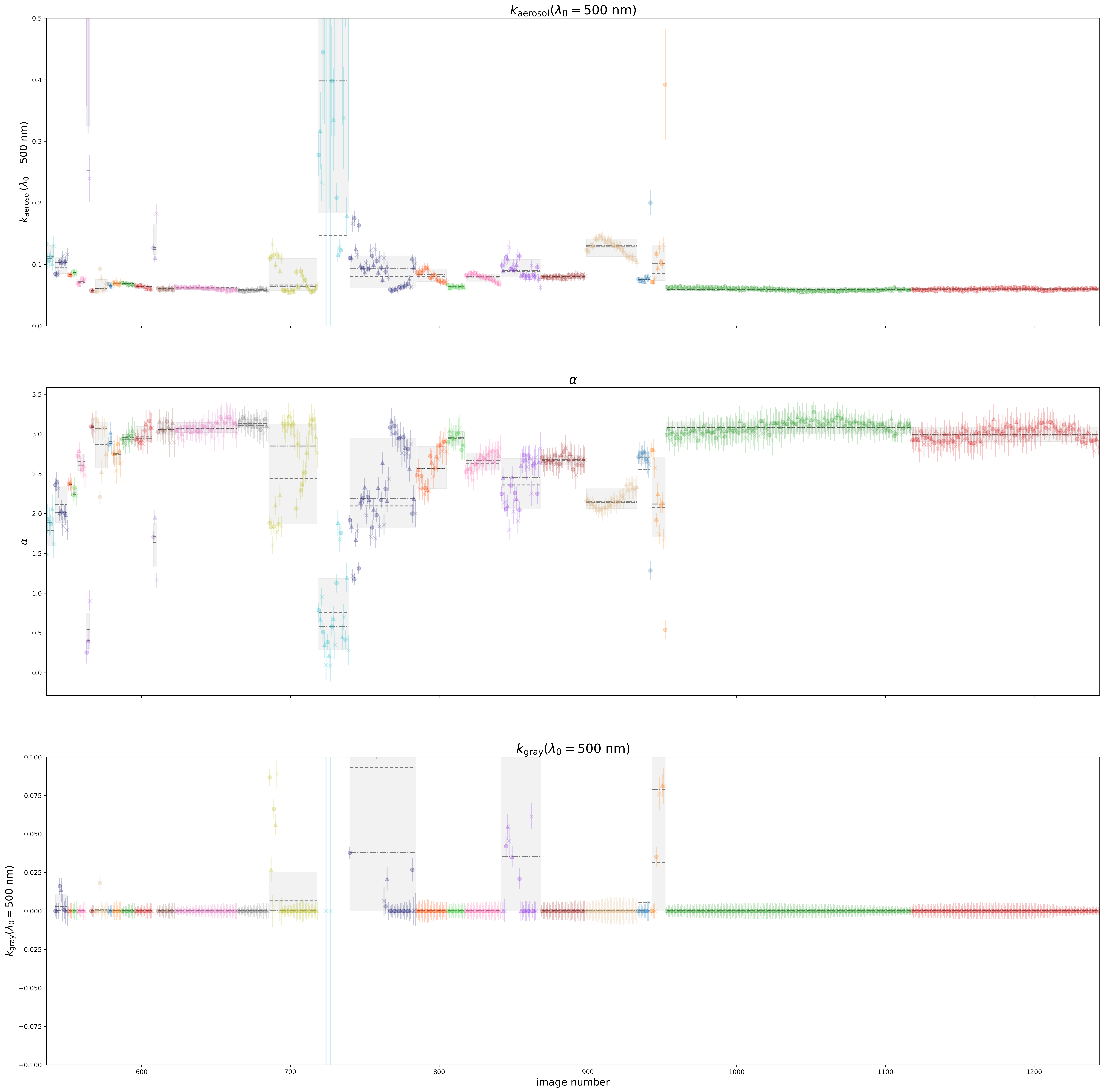}
    \caption{The fit parameters for per image for the Angstrom model plus a grey offset. First image show the amplitude of the Aerosol extinction $k_\text{Aerosol}(\lambda_0=500\text{nm})$, Each night is color coded separately with the 1 sigma range around, the mean value and the median value for each night also indicated. Second image is the $\alpha$ parameter, as above the nights are color coded, the 1 sigma ranges, the means, and medians are also indicated. Third image is the same as above, except here it is the $k_\text{gray}(\lambda_0=500\text{nm})$ which is the gray offset we allowed, to account for possible cloud contamination or other non wavelength dependent issues.}
    \label{fig:extincnight}
\end{figure}

\end{document}